\documentclass[11pt,a4paper]{article}
\pdfoutput=1
\usepackage{jheppub}

\usepackage{multirow}
\usepackage{amsmath,amssymb}
\usepackage{graphicx}
\usepackage{color}
\usepackage{hyperref}
\usepackage{url}
\usepackage{slashed}
\usepackage[usenames,dvipsnames]{xcolor}
\usepackage{amsmath}
\usepackage{graphicx}
\usepackage{subcaption}
\usepackage{amssymb}
\usepackage{hyperref}
\usepackage{amsthm}
\usepackage{MnSymbol}
\usepackage{nicefrac}
\usepackage{subcaption}
\usepackage{amsfonts}
\usepackage{dsfont}

\usepackage[markup=nocolor]{changes}
\usepackage{natbib}

\usepackage{graphicx}
\usepackage{epstopdf}

\usepackage[normalem]{ulem}

\usepackage{dcolumn}
\usepackage{bm}

\usepackage{amssymb}
\usepackage{slashed}
\usepackage{graphicx}
\usepackage{graphics}
\usepackage{epsfig}
\usepackage{color}
\usepackage{soul}
\usepackage{tabularx}
\usepackage{amsmath}
\usepackage{amsfonts}
\usepackage{amssymb}
\usepackage{hyperref}
\usepackage{microtype}

\newcommand{\bea}{\begin{eqnarray}}
\newcommand{\eea}{\end{eqnarray}}
\newcommand{\be}{\begin{equation}}
\newcommand{\ee}{\end{equation}}
\usepackage[normalem]{ulem}

\def\s0#1#2{\mbox{\small{$ \0{#1}{#2} $}}}
\def\0#1#2{\frac{#1}{#2}}

\title{Asymptotic safety in the dark}
\author[a]{Astrid Eichhorn,}
\author[a]{Aaron Held,}
\author[a,b]{Peter Vander Griend}
\emailAdd{a.eichhorn@thphys.uni-heidelberg.de, a.held@thphys.uni-heidelberg.de,vandergriend@tum.de}

\affiliation[a]{Institut f\"ur Theoretische Physik, Universit\"at Heidelberg, Philosophenweg 16, 69120 Heidelberg, Germany}
\affiliation[b]{Physik Department, Technische Universit\"at M\"unchen, D-85748 Garching, Germany}

\title{Asymptotic safety in the dark
}

\abstract{
We explore the Renormalization Group flow of massive uncharged fermions -- a candidate for dark matter -- coupled to a scalar field through a Higgs portal. We find that fermionic fluctuations can lower the bound on the scalar mass that arises from vacuum stability. Further, we discuss that
despite the perturbative nonrenormalizability of the model, it could be ultraviolet complete at an asymptotically safe fixed point. In our approximation, this simple model exhibits two mechanisms for asymptotic safety: a balance of fermionic and bosonic fluctuations generates a fixed point in the scalar self-interaction; asymptotic safety in the portal coupling is triggered through a balance of canonical scaling and quantum fluctuations.
As a consequence of asymptotic safety in the dark sector, the low-energy value of the portal coupling could become a function of the dark fermion mass and the scalar mass, thereby reducing the viable parameter space of the model.
}

\begin{document}
\maketitle

\section{Introduction}

Astrophysical observations provide clues that something is missing in our understanding of the degrees of freedom that make up the matter in our universe. Modifications of gravity as well as additional matter degrees of freedom are both considered as possible candidates. A particularly simple model contains an additional ``dark'', i.e., uncharged fermion $\psi$, that couples to the Higgs field $H$ of the Standard Model through a Higgs-portal interaction \cite{LopezHonorez:2012kv,Fedderke:2014wda} of the form
\be
\label{eq:hp-langrangian}
\mathcal{L}_{\rm HP} =  \bar{\lambda}_{h\psi} \bar{\psi}\psi\, H^{\dagger}H.
\ee 
At an appropriate interaction  strength, dark matter with an abundance in accordance with observations can be produced thermally in the early universe, i.e., $\psi$ is a thermal relic \cite{Bergstrom:2000pn,Bertone:2004pz}. Additionally, the Higgs portal opens the door for direct searches at dedicated experiments, e.g., \cite{Akerib:2016vxi,Aprile:2016swn} as well as at the LHC, see, e.g., \cite{Kanemura:2010sh,Djouadi:2011aa,Djouadi:2012zc,Hoferichter:2017olk}. These have already succeeded in excluding a part of the parameter space spanned by the fermion mass $M_{\psi}$ and the Higgs-portal coupling $\bar\lambda_{h\psi}$, see, e.g., \cite{Beniwal:2015sdl}.
\\
In the effective field theory setting, the mass and the portal coupling are independent free parameters. Demanding that $\psi$ constitutes the complete dark matter relic abundance imposes a constraint on the cross-section of Higgs-fermion scattering processes, translating into a line in the $M_{\psi}-\bar{\lambda}_{h\psi}$ plane. This still leaves a huge range of masses for experiments to explore, making it difficult to completely exclude this model as a viable dark-matter model, even under the very strong assumption that no additional dark matter degrees of freedom exist. Our first aim in this paper is to highlight that if the model is analyzed from an asymptotically safe point of view an additional constraint linking $M_{\psi}$ to $\bar{\lambda}_{h\psi}$ might arise from the consistency of the underlying ultraviolet (UV) physics. 
At the same time, such a setting addresses the problem that the Higgs-portal coupling has negative mass dimension and the model is therefore not perturbatively renormalizable.
The asymptotic safety scenario goes beyond the paradigm of perturbative renormalizability.  It can provide a UV completion for effective field theories, even if they feature higher-order couplings.
In this case, the running couplings approach finite values at high momenta,  triggered by an underlying, interacting Renormalization Group (RG) fixed point.  Such a fixed point can also be achieved for perturbatively nonrenormalizable couplings.
These higher-order couplings cannot become asymptotically free, as their negative scaling dimension implies that they cannot deviate from zero, once they have been set to zero in the UV.
In the asymptotically safe case, residual interactions are present which alter the scaling dimensions of couplings and can generate an interacting fixed point at high energies.\\
For instance, a particularly intriguing case is that of gravity. Despite its perturbative nonrenormalizability, a quantum field theory of the metric could exist within the paradigm of asymptotic safety. Strong indications in favor of the theoretical viability of such a model have been found, see, e.g., \cite{Weinberg:1980gg,Reuter:1996cp, Benedetti:2009rx, Falls:2013bv, Gies:2016con, Denz:2016qks}, and promising hints suggest the possibility of a predictive asymptotically safe model of quantum gravity and Standard Model matter, see \cite{Shaposhnikov:2009pv, Harst:2011zx, Eichhorn:2017ylw, Eichhorn:2017egq, Wetterich:2016uxm,Eichhorn:2017lry, Eichhorn:2017als, Gies:2018jnv} and references therein. In that case, the search for asymptotic safety is complicated by conceptual and technical challenges. Thus, finding a simpler candidate for an asymptotically safe model in $d=4$ to analyze the paradigm's implications is desirable.
\\
Interacting RG fixed points are well known in statistical physics, defining universality classes for continuous phase transitions. Typically, the setting of these models is that of three (Euclidean) dimensions. In high-energy physics, asymptotic safety was recently discovered in perturbation theory. It was found that gauged Yukawa systems in four dimensions, i.e., models containing a large number of scalars, fermions and vector bosons, can become asymptotically safe \cite{Litim:2014uca,Bond:2017wut,Mann:2017wzh,Pelaggi:2017abg}.
\\
Here we aim at providing indications for asymptotic safety in a model in $d=4$ that is very simple in terms of the degrees of freedom it contains: We consider the Higgs portal model with fermions and scalars and find hints for asymptotic safety in a truncated system of RG equations. In particular, the system is intriguing as it exhibits two mechanisms that can lead to asymptotic safety: Firstly, the Higgs-portal coupling with negative mass dimension reaches an asymptotically safe regime, as canonical scaling and quantum scaling balance. Secondly, the quartic Higgs coupling which has vanishing canonical dimension becomes asymptotically safe because fermionic and bosonic fluctuations balance. The latter mechanism plays a role in the case of asymptotically safe gauged Yukawa systems, while the former is 
key for quantum gravity. The Higgs-portal fermion model could be a particularly simple case to exemplify these mechanisms. 

Further motivation to study the RG flow of the model stems from the compatibility of the observed Higgs mass with vacuum stability: Within the Standard Model, the Higgs quartic coupling must be negative at the Planck scale -- at least for the central value of the measured top mass -- in order to accommodate a Higgs mass of 125 GeV \cite{Maiani:1977cg,Altarelli:1994rb,Casas:1994qy,Casas:1996aq,Schrempp:1996fb, Isidori:2001bm, Isidori:2007vm, Ellis:2009tp, EliasMiro:2011aa, Degrassi:2012ry,Bezrukov:2012sa, Buttazzo:2013uya,Gies:2013fua,Eichhorn:2015kea,Borchardt:2016xju}. Additional degrees of freedom that couple to the Higgs can impact the running of its potential and thereby either increase or reduce the tension between a vanishing quartic Higgs coupling at the Planck scale and the measured Higgs mass.  
\\
This paper is structured as follows: We introduce the key ideas underlying asymptotic safety in Sec.~\ref{sec:as_and_pred}, derive the beta functions in a truncation of the full dynamics with functional RG techniques in Sec.~\ref{sec:FRG}, analyze the RG flow and fixed-point properties in Secs.~\ref{sec:HPRGflow} and \ref{sec:results}, respectively, and conclude in Sec.~\ref{sec:conclusions}. 
\section{Asymptotic safety and predictivity}
\label{sec:as_and_pred}

An asymptotically safe UV completion is defined by an RG fixed point at which the couplings take finite values. Specifically, the fixed point exists for the dimensionless versions of all couplings, i.e., in the fixed-point regime all dimensionful quantities scale according to their canonical dimension. Finiteness in the dimensionless couplings suffices to guarantee finiteness of observables, such as, e.g., scattering cross-sections, at all energy scales \cite{Weinberg:1980gg}. Asymptotic safety provides a generalization of asymptotic freedom, where the UV fixed point is a free fixed point. In contrast, theories exhibiting asymptotic safety feature a scale invariant, but interacting rather than free, regime in the UV. Towards the infrared (IR), the RG flow drives the couplings away from their fixed-point values. The deviation is encoded in a set of free parameters, corresponding to the relevant couplings.
\\
For a theory with couplings \(g_{i}\) and corresponding beta functions $\beta_{g_{i}}(g_{j}) =\partial_t g_i(k)= k\partial_k\, g_i(k)$ asymptotic safety entails the condition
\begin{equation}
	\beta_{g_{i}}(g_{j})=0. 
\end{equation}
For a given model governed by a specific set of symmetries, in general an infinite number of operators obey those symmetries; the couplings $g_i$ corresponding to these operators span the infinite dimensional theory space. To analyze the predictivity of the model, let us define the stability matrix \(\mathrm{B}_{ij}\) describing the linearized flow in the vicinity of a fixed point at $g_i=g_i^{\ast}$ as
\begin{equation}\label{eq:stabmatdef}
	B_{ij} = \frac{\partial  \beta_{g_i}}{\partial g_j} \Bigg|_{\vec{g}= \vec{g}^{\ast}}.
\end{equation}
The eigenvalues of the stability matrix multiplied by an additional negative sign,
\be
\theta_I= -{\rm eig}\, B_{ij},
\ee
are the critical exponents. To see the connection between the critical exponents and predictivity, we note that the definition of the stability matrix in Eq.~\eqref{eq:stabmatdef} is also the leading- order term in the Taylor expansion of the beta functions about the fixed point. The solution to the linearized RG flow provides the couplings, at linear approximation, in the vicinity of the fixed point
\begin{equation}
	g_{i}(t)=g_{i}^{*}+\sum_I c_I\, V_i^{I}\, e^{-\theta_I\, t},
\end{equation} 
where the \(V_{I}\) are the eigenvectors of the stability matrix corresponding to the eigenvalues \(-\theta_{I}\) and the \(c_{I}\) are a priori undetermined constants of integration. 
However, only a subset of those actually plays a role at low energies, allowing the model to be 
predictive.
The directions with negative critical exponents are called irrelevant. Along those directions, the RG flow towards the IR pulls the couplings towards the fixed-point value, irrespective of the value of $c_I$. Therefore, the corresponding $c_I$ is irrelevant. 
In contrast, the positive critical exponents and corresponding directions are relevant parameters and directions. Deviations from the fixed point along a relevant direction determine the low-energy physics: The RG flow to the IR deviates from the fixed point on a trajectory within the critical hypersurface spanned by the relevant directions. A free parameter -- indicated by the corresponding $c_I$ -- is associated to each such direction. The values of the relevant parameters must be taken as input from experiment; demanding asymptotic safety then  determines the values of the irrelevant couplings as functions of the relevant couplings.
\\
A model is nonperturbatively renormalizable if it is defined on an RG trajectory that emanates from an asymptotically safe fixed point with a finite number of relevant directions.
Crucially, this can be realized also for perturbatively nonrenormalizable theories with couplings of negative mass dimension.

\section{Derivation of beta functions}\label{sec:FRG}
\subsection{Functional Renormalization Group}
The functional RG is based on Wilson's  concept of integrating out momentum shells in the Euclidean generating functional. For the effective action, this is achieved by augmenting the generating functional by the addition of a mass-like, RG-scale-dependent and momentum-dependent cutoff term that suppresses low-momentum quantum fluctuations. Lowering the RG scale $k$ successively allows to take into account progressively more IR field configurations.
Taking a (modified) Legendre transform provides the flowing action $\Gamma_k[\Phi]$, which depends on the expectation value of the fields in the model, summarized in the ``superfield" $\Phi$. $\Gamma_k$ smoothly interpolates between the microscopic action in the UV and the standard effective action $\Gamma_{k\rightarrow0}[\Phi]= \Gamma[\Phi]$ in the IR. \(\Gamma_k[\Phi]\) contains all operators and corresponding couplings $g_i(k)$ obeying the symmetries of the theory. 
The scale-dependence of couplings can be read off from the scale dependence of $\Gamma_k[\Phi]$,
\be
\partial_t \Gamma_k[\Phi] = \sum_i \beta_{g_i}\int d^dx\, \mathcal{O}_i,
\ee
by projecting the full flow $\partial_t \Gamma_k[\Phi]$ onto the field monomial $\mathcal{O}_i$ corresponding to the coupling $g_i (k)$ of interest. The functional RG equation, also known as the Wetterich equation \cite{Wetterich:1992yh, Morris:1993qb}, encodes the flow in theory space in 
a functional differential equation
\begin{equation}
	\label{eq:flowEq}
	\partial_{t}\Gamma_{k}[\phi]=\frac{1}{2}\mathrm{STr}\, \partial_t R_k \left( \Gamma^{(2)}_{k}+R_{k}\right)^{-1},
\end{equation}
where STr represents a supertrace summing over all discrete indices and integrating over all continuous indices. For Grassmann-valued fields, such as the dark fermion, an additional negative sign is included. $\Gamma_k^{(2)} = \overset{\rightarrow}{\frac{\delta}{\delta\Phi_i^T}}\Gamma_k  \overset{\leftarrow}{\frac{\delta}{ \delta\Phi_j}}$ is a matrix in field-space. 
In our case, $\Phi^T=\left(\phi, \psi^T, \bar{\psi}\right)$, where the fields can also carry additional internal indices. 
The regulator  ensures IR and UV regularity in the flow equation. This is a direct consequence of the fact, that an IR regulator, inserted as a mass-like term in the generating functional should satisfy $R_k(p^2) >0 $ for $k^2>p^2$ (suppression of IR modes with momenta $p^2<k^2$) and $R_k(p^2) \rightarrow 0$ for $k^2<p^2$ (allows UV modes to be integrated out) and thus $\partial_t R_k(p^2)\rightarrow 0$ for $p^2>k^2$.
Eq.~\eqref{eq:flowEq} gives, in one-loop-exact form, the running of the flowing action of a theory and will be the main tool utilized in our analysis in the rest of this paper. Once a basis in theory space is chosen, e.g., by performing a derivative expansion, the flow equation translates into an infinite tower of coupled differential equations for the corresponding beta functions. For practical purposes, it is necessary to resort to an approximation, i.e., a truncation of theory space is chosen. Physically, this corresponds to neglecting some part of the dynamics. A good choice of truncation that neglects only sub-leading parts of the dynamics is crucial to obtain robust results.
\\
For reviews on the functional RG, see, e.g., \cite{Berges:2000ew, Polonyi:2001se,
Pawlowski:2005xe, Gies:2006wv, Delamotte:2007pf, Rosten:2010vm, Braun:2011pp}.

\subsection{Beta functions for the Higgs portal model}
We study the Euclidean action of a scalar $O(N)$-model coupled to $N_f$ fermionic degrees of freedom with a $U(N_f)$ symmetry for the fermions which reduces to a $\mathbb{Z}_2$ symmetry for one flavor, i.e., $N_f=1$. This includes a complex scalar field for $N=2$, and four degrees of freedom, as the Standard Model Higgs, for $N=4$.
The flowing action $\Gamma_k$ contains all terms compatible with these symmetries. 
This excludes a Yukawa-interaction between fermions and scalars.
We employ a derivative expansion and focus on the (next-to-) leading order. Furthermore, we include only terms up to quadratic order in the fermions, neglecting, e.g., four-fermion interactions, which carry canonical dimension -2. Thus our truncation reads
\be
\label{eq:effectiveAction}
	\Gamma_k = \int d^dx \Big[
		i\sum_{j=1}^{N_f}\bar{\psi}^j(Z_\psi\slashed\partial
		)\psi^j + \frac{1}{2}Z_\phi\sum_{a=1}^N\partial_\mu \phi^a \partial^\mu \phi^a
		+ i
		\sum_{j=1}^{N_f}\bar\psi^j\psi^j
		\bar{V}(\rho) + \bar{U}(\rho)
	\Big]\;,
\ee
where $\rho = \frac{1}{2}\sum_{a=1}^N\phi^a\phi^a$. $Z_{\phi}=Z_{\phi}(k)$ and $Z_{\psi}=Z_{\psi}(k)$ are scale-dependent wave-function renormalizations for the fields related to the anomalous dimensions by
\begin{align}
	\eta_\phi = -\partial_t\ln Z_\phi\;,\quad
	\eta_\psi = -\partial_t\ln Z_\psi\;.
\end{align} 
We carry out our analysis in two regimes of the scalar potential
\begin{align}
	\label{eq:potentials}
	\bar{U}_\text{SYM}(\rho) &= \bar{m}_\phi^2 \rho + \sum_{n=2}^{N_t} \frac{\bar\lambda_{2n}}{n!}\rho^n,
	\quad\quad\quad
	\bar{U}_\text{SSB}(\rho) = \sum_{n=2}^{N_t} \frac{\bar\lambda_{2n}}{n!}(\rho-\bar\kappa)^n
	\;,
\end{align}
where $N_t$ is the maximum order of the truncation, and the shorthand $\rm LPA_{N_t}^{(')}$ is used to denote such truncations, with the prime indicating the inclusion of the anomalous dimensions. We explicitly explore $N_t=1,...,20$.
Furthermore, we specify the portal interactions to be
\begin{align}
	\bar{V}_\text{SYM}(\rho) &= \bar{m}_\psi + \bar\lambda_{h\psi}\rho,
	\quad\quad\quad
	\bar{V}_\text{SSB}(\rho) = \bar{m}_\psi + \bar\lambda_{h\psi}(\rho-\bar\kappa)
	\;.
\end{align}
The potential is expanded around its minimum in the symmetric and symmetry-broken phase with $\bar{\kappa}$ denoting a nontrivial vacuum expectation value for the scalar in the regime of spontaneous symmetry breaking (SSB).
\\
For our calculations, we employ bosonic, \(R_{k}^{B}\), and fermionic, \(R_{k}^{F}\), Litim-type regulators \cite{Litim:2000ci}:
\begin{align}
	R_{k}^{B}&=Z_{\phi}\left( k^{2}-p^{2} \right)\theta \left( k^{2}-p^{2} \right),\\
	R_{k}^{F}&= Z_{\psi}\slashed{p} \left(\sqrt{\frac{k^2}{p^2}}-1 \right)  \theta \left( k^{2}-p^{2} \right)\;.
\end{align}
To explicitly evaluate the beta functions, we perform an expansion of $\Gamma_k^{(2)}$ into terms with and without fermionic field content. This allows us to expand the right-hand-side of Eq.~\eqref{eq:flowEq} up to second order in the fermionic fields, which is all we need in our truncation. Specifically, 
\be
\Gamma_k^{(2)}+R_k= \mathcal{P}_k +\mathcal{F}_k,
\ee
where $\mathcal{P}_k=\left(\Gamma_k^{(2)}+R_k\right)|_{\psi=0=\bar{\psi}}$, and $\mathcal{F}_k$ is the corresponding remainder term. Together with the definition of $\tilde{\partial}_t$, which only acts on the $k$-dependence in $R_k$, this allows to rewrite Eq.~\eqref{eq:flowEq},
\be
\partial_t \Gamma_k= \frac{1}{2}{\rm STr}\tilde{\partial}_t\mathcal{P}_k^{-1} + \frac{1}{2}\sum_n \frac{(-1)^{n-1}}{n}{\rm STr}\tilde{\partial}_t \left(\mathcal{P}_k^{-1}\mathcal{F}\right)^n.
\ee
In our case, terms in $\mathcal{F}$ are either linear or quadratic in the fermion fields, and thus we can terminate the expansion at $n=2$. Explicitly,
\begin{align}
\mathcal{P}_k(p,q) &=
\begin{pmatrix}
	\delta^{ab}\!\left(Z_\phi p^2+\bar{U}'(\rho)\!\right)+\phi^a\phi^b \bar{U}''(\rho) & 0 & 0 
	\\
	0 & 0 & (-Z_\psi\slashed{p}^T-i\bar{V}(\rho))\delta^{ij}
	\\
	0 & (-Z_\psi\slashed{p}+i\bar{V}(\rho))\delta^{ij} & 0
\end{pmatrix}
\delta_{p,q},
\\
\mathcal{F}_k(p,q) &=
\begin{pmatrix}
	i\bar{\psi}^l\psi^l\left(\delta^{ab}\bar{V}'(\rho)+\phi^a\phi^b\bar{V}''(\rho)\right)
	& i\bar\psi^j\phi^b\bar{V}'(\rho) & -i\psi^{Tj}\phi^b\bar{V}'(\rho) 
	\\
	-i\bar\psi^{Ti}\phi^a\bar{V}'(\rho) & 0 & 0
	\\
	i\psi^i\phi^a\bar{V}'(\rho) & 0 & 0
\end{pmatrix}
\delta_{p,q}\;.
\end{align}
The right-hand side of the Wetterich equation can then be evaluated by performing the matrix multiplication, taking the trace over the Dirac indices, including an additional negative sign for fermions and performing the momentum integral. 
To isolate the running of specific couplings from 
the Wetterich equation, we project onto the corresponding field monomials. For example,
\be
\partial_t \bar{U}_k =\frac{1}{\Omega} \partial_t \Gamma_k\Big|_{\psi=0=\bar{\psi}; \rho=\rm const},
\ee
where $\Omega$ is the spacetime volume. In particular, suitable projections, cf. Eq.~\eqref{eq:projections} in App.~\ref{sec:projectionRules}, give the running of the dimensionless fermionic couplings $m_\psi$ and $\lambda_{h\psi}$. We state these general $\beta$-functions in Eq.~\eqref{eq:runningCouplingsAndPotential} in App.~\ref{sec:explicitGeneralBetas}.
The transition from dimensionful (with bars) to dimensionless couplings (without bars) is given by 
\begin{align}
\label{eq:dimToDimless}
	m_\psi &= \frac{\bar{m}_\psi}{Z_\psi k}\;,\quad
	m_\phi = \frac{\bar{m}_\phi}{Z_\phi^{1/2} k}\;,\quad
	\kappa = \frac{\bar{\kappa}}{Z_\phi k^2},
	\notag\\
	\lambda_{h\psi} &= \frac{\bar\lambda_{h\psi}}{Z_\psi Z_\phi k^{3-d}},\;\quad
	\lambda_{2i} = \frac{\bar\lambda_{2i}}{Z_\phi^i\,k^{d-(d-2)^i}}\;.
\end{align}
This results in the following expressions
\begin{align}
	\partial_t U(\rho)  =& \left(d-2+\eta_\phi\right)U'(\rho)\rho - d\,U(\rho) 
	\\\notag&
	+ l_0^{(B)d}(\omega_\rho(\rho);\eta_\phi) 
	+ (N -1)\,l_0^{(B)d}(\omega_G(\rho);\eta_\phi) 
	- N_f\,d_\gamma\,l_0^{(F)d}(\omega_\psi(\rho);\eta_\psi)\;,
	\\[1em]
	\partial_t m_\psi  =& \left(-1+\eta_\psi\right)m_\psi+\lambda_{h\psi}\partial_t\kappa + (d-2+	\eta_\phi)\lambda_{h\psi}\kappa	
	\notag\\&
	-\lambda_{h\psi}\left(l_1^{(B)d}(\omega_\rho;\eta_\phi) + (N -1)\,l_1^{(B)d}(\omega_G;\eta_	\phi)\right)
	\notag\\&
	+4\kappa\lambda_{h\psi}^2\sqrt{\omega_\psi}\,
	l_{1,1}^{(BF)d}(\omega_\rho,\omega_\psi;\eta_\phi,\eta_\psi)
	\;,\\[1em]
	\partial_t \lambda_{h\psi}  =& 
	\left(d-3+\eta_{\psi}+\eta_{\phi}\right) \lambda_{h\psi}
	\notag\\&
	+ 2\lambda_{h\psi}\,
	\omega_\rho'\left(l_0^{(B)d}(\omega_\rho;\eta_\phi) + (N -1)\omega_G'\,l_0^{(B)d}(\omega_G;\eta_\phi)\right)
	\notag\\&
	+4\left( \kappa\lambda_{h\psi}^3 + \lambda_{h\psi}^2\sqrt{\omega_\psi} \right)
	l_{1,1}^{(BF)d}(\omega_\rho,\omega_\psi;\eta_\phi,\eta_\psi)
	\notag\\&
	-8\kappa\lambda_{h\psi}^3\omega_\psi\,
	l_{1,2}^{(BF)d}(\omega_\rho,\omega_\psi;\eta_\phi,\eta_\psi)
	\notag\\&
	-4\kappa\lambda_{h\psi}^2\sqrt{\omega_\psi}\,\omega_\rho'\,
	l_{2,1}^{(BF)d}(\omega_\rho,\omega_\psi;\eta_\phi,\eta_\psi)
	\;.
\end{align}
In our truncation, 
\be
\eta_{\psi}=0,\quad \eta_{\phi} =
\frac{\kappa(3\lambda_4 + 2\kappa\lambda_6)^2}{16\pi^2(1+2\kappa\lambda_4)^4},
\ee
where the explicit threshold integrals $l_i^d$ integrated with Litim-type cutoff function are given in App.~\ref{app:thresholdIntegrals}. The effective squared masses of the radial mode $\omega_\rho$, the Goldstone mode $\omega_G$ and the fermions $\omega_\psi$ are given by
\begin{align}
	\omega_\rho &= U'(\rho) + 2\rho U''(\rho)
	\;,\quad
	\omega_G = U'(\rho)
	\;,\quad
	\omega_\psi = 
	V(\rho)^2
	\;.
	\label{eq:effectiveMasses}
\end{align}
We explicitly confirm that in the corresponding limits, our results are in agreement with those for the flow of the scalar potential \cite{Litim:2002cf} and for the Higgs-portal coupling \cite{Vacca:2015nta}.
\section{Renormalization Group flow in the Higgs portal to fermionic dark matter \& Higgs-mass bound}\label{sec:HPRGflow}
We now analyze the RG flow in the Higgs-portal model.
To be specific, we initially focus on $N=1,\, N_f=1$ and expand to quartic order in the symmetric scalar potential. Furthermore, we begin by setting $\eta_{\phi}=0=\eta_{\psi}$.
The beta functions in this truncation are given by
\begin{align}
\label{eq:beta_portal_SYM}
	\beta_{\lambda_{h\psi}} &=
	\lambda_{h\psi}
	+\frac{3\lambda_4\,\lambda_{h\psi}}{16\pi^2(1+m_\phi^2)^3}
	+\frac{m_\psi\,\lambda_{h\psi}^2(2+m_\psi^2+m_\phi^2)}{8\pi^2(1+m_\phi^2)^2(1+m_\psi^2)^2}\;,
	\\
	\label{eq:beta_quartic_SYM}
	\beta_{\lambda_4} &=
	 \frac{9\lambda_4^2}{16\pi^2(1+m_\phi^2)^3}
	+\frac{\lambda_{h\psi}^2}{4\pi^2(1+m_\psi^2)^2}
	- \frac{m_\psi^2\,\lambda_{h\psi}^2}{\pi^2(1+m_\psi^2)^3}\;,
	\\
	\label{eq:beta_mPsi_SYM}
	\beta_{m_\psi} &= 
	-m_\psi 
	- \frac{\lambda_{h\psi}}{32\pi^2(1+m_\phi^2)^2}\;,
	\\
	\label{eq:beta_mPhi_SYM}
	\beta_{m_\phi^2} &= -2m_\phi^2
	+ \frac{m_\psi\lambda_{h\psi}}{4\pi^2(1+m_\psi^2)^2}
	-\frac{3\lambda_4}{32\pi^2(1+m_\phi^2)^2}\;,
\end{align}
where the nontrivial denominators are a consequence of the Functional RG setup. They lead to automatic decoupling of massive degrees of freedom, once the RG scale $k$ drops below the corresponding mass. Setting those threshold contributions to one reproduces the universal one-loop coefficient of the beta function for the dimensionless quartic coupling $\lambda_4$.

In particular, it is of interest to explore the impact of fermionic fluctuations on the Higgs potential. Fermions which are coupled to the Higgs through a standard Yukawa interaction of the form $\bar{\psi}\psi H\, +\, h.c.$ lead to a lower bound on the Higgs mass if the condition of vacuum stability is imposed \cite{Maiani:1977cg,Altarelli:1994rb,Casas:1994qy,Casas:1996aq,Schrempp:1996fb, Isidori:2001bm, Isidori:2007vm, Ellis:2009tp, EliasMiro:2011aa, Degrassi:2012ry,Bezrukov:2012sa, Buttazzo:2013uya,Gies:2013fua,Eichhorn:2015kea,Borchardt:2016xju}: This is a consequence of a \emph{negative} fermionic contribution to the beta function for the Higgs quartic coupling. Starting with a vanishing Higgs quartic coupling at an UV scale $\Lambda$, this term leads to a growth of $\lambda_4$ towards the infrared. The larger the fermion mass, i.e., the Yukawa coupling, and the larger the UV scale $\Lambda$, the larger the resulting IR value of the Higgs quartic coupling $\lambda_4$. As this sets the mass of the Higgs, low Higgs masses can only be reached if the initial UV value of $\lambda_4$ is chosen negative, typically interpreted as indication for an unstable Higgs potential at that scale. \\
Intriguingly, the situation differs for a fermion coupled through a Higgs-portal interaction. Depending on the value of the fermion mass, the fermionic fluctuations can add a positive or a negative contribution to $\beta_{\lambda_4}$, cf.~Eq.~\eqref{eq:beta_quartic_SYM}. For the dimensionless mass $m_{\psi}^2<1/3$, fermionic fluctuations yield a positive contribution to the running of the Higgs quartic coupling. This is structurally identical to the Higgs portal to scalar dark matter.
It is well-known that scalar dark matter coupled through a Higgs portal leads to a stabilization of the Higgs potential, in other words, it lowers the lower bound on the Higgs mass \cite{Gonderinger:2009jp,Clark:2009dc,Lerner:2009xg,Gonderinger:2012rd}. \\
On the other hand $m_{\psi}^2>1/3$ corresponds to a regime where fermionic fluctuations coupled through a Higgs portal act in a similar way as those coupled through Yukawa interactions, leading to an increase of the quartic coupling towards the infrared. Accordingly, in this regime the lower bound on the Higgs mass arising from the demand of vacuum stability would be shifted towards \emph{larger} values.\\
Sufficiently light fermionic dark matter accordingly contributes to lowering the lower bound on the Higgs mass. As the corresponding fermionic contributions to $\beta_{\lambda_4}$ should also be present for the Higgs quartic coupling in the Standard Model, a similar effect is to be expected. Therefore, fermionic dark matter might contribute to reconciling the measured Higgs mass of 125 GeV with a vanishing (instead of slightly negative) Higgs quartic coupling at the Planck scale.
\section{Asymptotic safety in the Higgs-portal to fermionic dark matter}
\label{sec:results}
\subsection{Mechanisms for asymptotic safety}
\label{sec:higgs-portal}
Asymptotic safety can arise through the balance of loop terms, e.g., through the balance of one-loop versus two-loop terms in a perturbative regime, see, e.g., \cite{Litim:2014uca, Bond:2018oco} or through the balance of fermionic versus bosonic loops at the same loop-order. The latter can also be present beyond the perturbative regime, i.e., the nonperturbative contributions from fermions and bosons can cancel to generate asymptotic safety, see, e.g., \cite{Gies:2009hq,Gies:2009sv,Vacca:2015nta} for corresponding studies.
A second possibility for asymptotic safety is present for dimensionful couplings, where canonical scaling can balance against quantum scaling, i.e., against loop effects, see, e.g., \cite{Gies:2003ic,Braun:2010tt}. The present model features both mechanisms. 

The Higgs-portal coupling has negative mass dimension and therefore features a canonical linear term in its beta function.
Quantum fluctuations can balance this dimensional running, schematically
\begin{align}
	\beta_{g_i} = \text{canonical}\times g_i - \text{quantum}\times g_i^2 = 0\;.
\end{align}
In general the quantum contribution can arise from a combination of different couplings in the model.
For the Higgs-portal coupling, cf.~Eq.~\eqref{eq:beta_portal_SYM}, this mechanism can be realized whenever the portal coupling and the dark-matter mass have opposite sign. In this case, the classical scaling $\sim\lambda_{h\psi}$ can be balanced by quantum fluctuations proportional to $\sim m_\psi\lambda_{h\psi}^2$.
A similar mechanism generating an interacting fixed point appears to be at work for the Newton coupling in asymptotically safe quantum gravity in $d>2$ dimensions \cite{Gastmans:1977ad,Christensen:1978sc,Falls:2015qga}, for the Wilson-Fisher fixed point in three dimensions \cite{Wilson:1971dc}, as well as for Yang-Mills theory in $d=4+\epsilon$ dimensions \cite{Peskin:1980ay,Gies:2003ic}.
This mechanism becomes accessible in perturbation theory in the vicinity of the critical dimension of an interaction, i.e., in $d=d_{\rm crit}\pm \epsilon$ dimensions, where the dimensional term is $\sim \epsilon$, and can balance against the one-loop term at an interacting fixed point of order $\epsilon$.\\

For couplings with vanishing mass dimension, i.e., perturbatively renormalizable ones, 
the dimensional running is absent and thus asymptotic safety can only be realized if different quantum fluctuations balance against each other. Typically, bosonic and fermionic fluctuations come with opposite sign and therefore models with both bosons and fermions are prime candidates to exhibit such a mechanism for interacting fixed points, schematically, see, e.g., \cite{Gies:2009hq}
\begin{align}
	\beta_{g_i} = \text{bosonic fluct.}\times g_b^2 - \text{fermionic fluct.} \times g_{f}^2\;.
\end{align}
The different contributions can in general arise from different interactions of the model, schematically indicated by a bosonic coupling $g_b$ and a fermionic coupling $g_f$.
In the beta function of the quartic scalar coupling, the contribution from fermionic fluctuations proportional to the Higgs-portal coupling can balance against the bosonic fluctuations proportional to the quartic coupling itself, cf. Eq.~\eqref{eq:beta_quartic_SYM}. In particular, the fluctuations of massive fermions can add effective antiscreening contributions balancing the screening effect of bosonic quantum fluctuations.
Here, the mechanism requires a non-vanishing portal coupling $\lambda_{h\psi}$ and a large enough dark matter mass $m_\psi$.  Note, that a fermion mass is expected to be present in the model as the Higgs-portal coupling breaks chiral symmetry.
\\
A similar mechanism is present in the conjectured quantum-gravity induced UV-completion for the Standard Model \cite{Eichhorn:2017egq}. In that case, matter and gravity fluctuations balance to induce an interacting fixed point which has a higher predictive power than the Standard Model itself. 
A related mechanism, with one-loop and two-loop terms cancelling against each other underlies the perturbative fixed points found in gauge-Yukawa systems with a large number of fields \cite{Litim:2014uca}.

\subsection{Fixed-point properties}

\begin{figure}
	\includegraphics[width=0.49\textwidth]{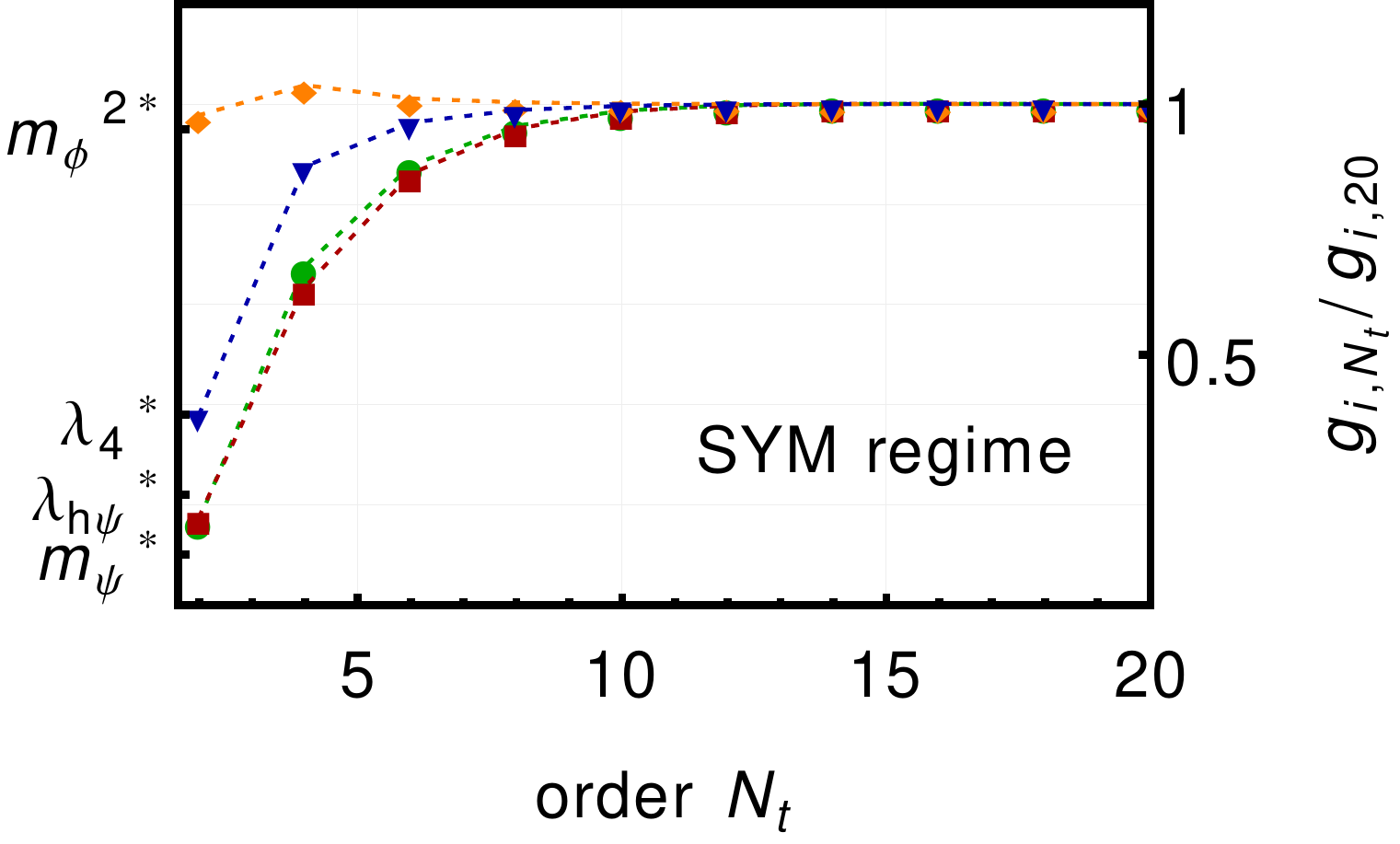}
	\hfill
	\includegraphics[width=0.49\textwidth]{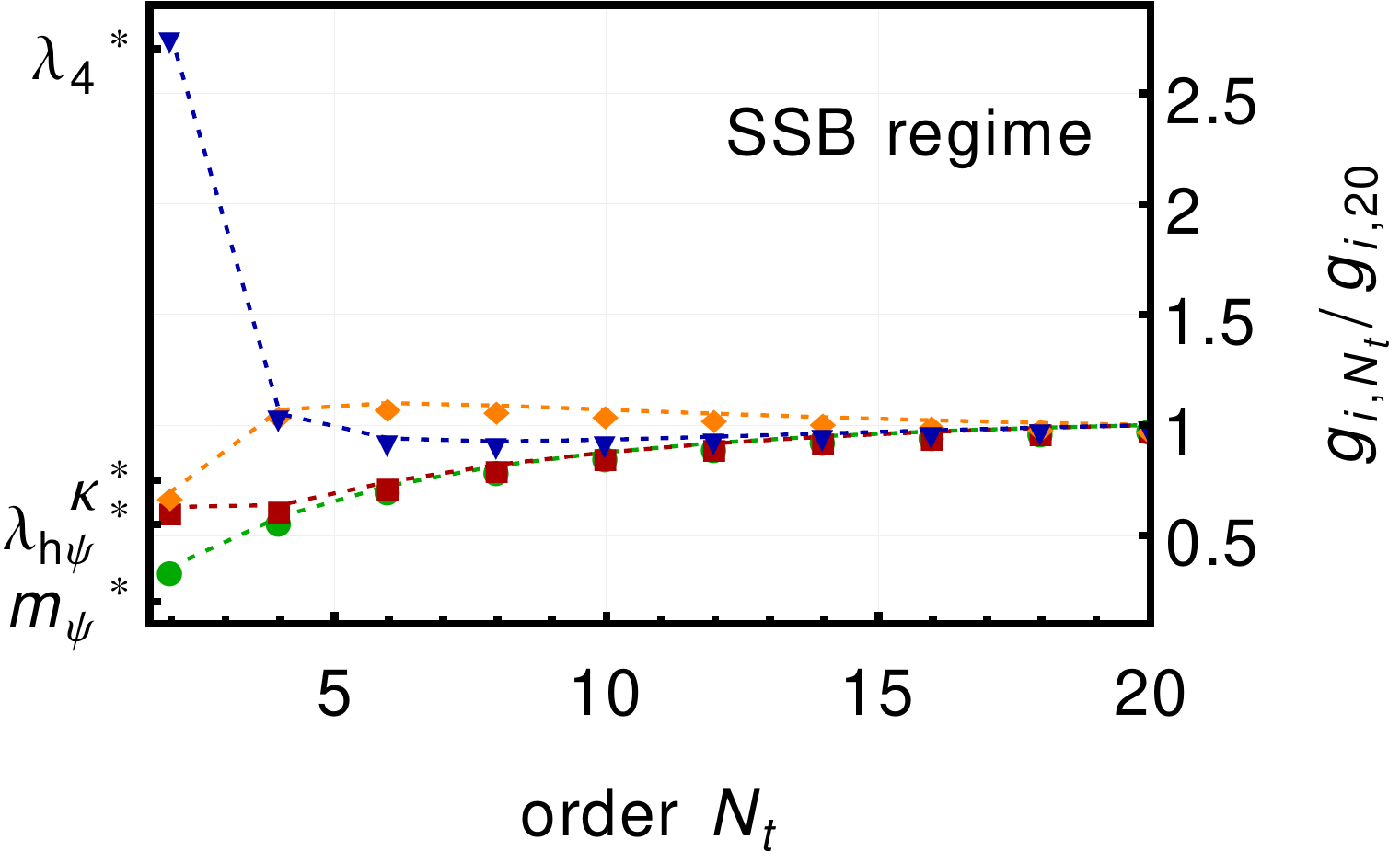}
\caption{
\label{fig:convergenceFPs}
Fixed point values for $N_f=N=1$ with increasing truncation order $N_t$ in the symmetric regime (left-hand panel) and in the regime of spontaneously broken symmetry (right-hand panel). All fixed point values are normalized to the values at the highest order.
}
\end{figure}
\begin{figure}
	\includegraphics[width=0.52\textwidth]{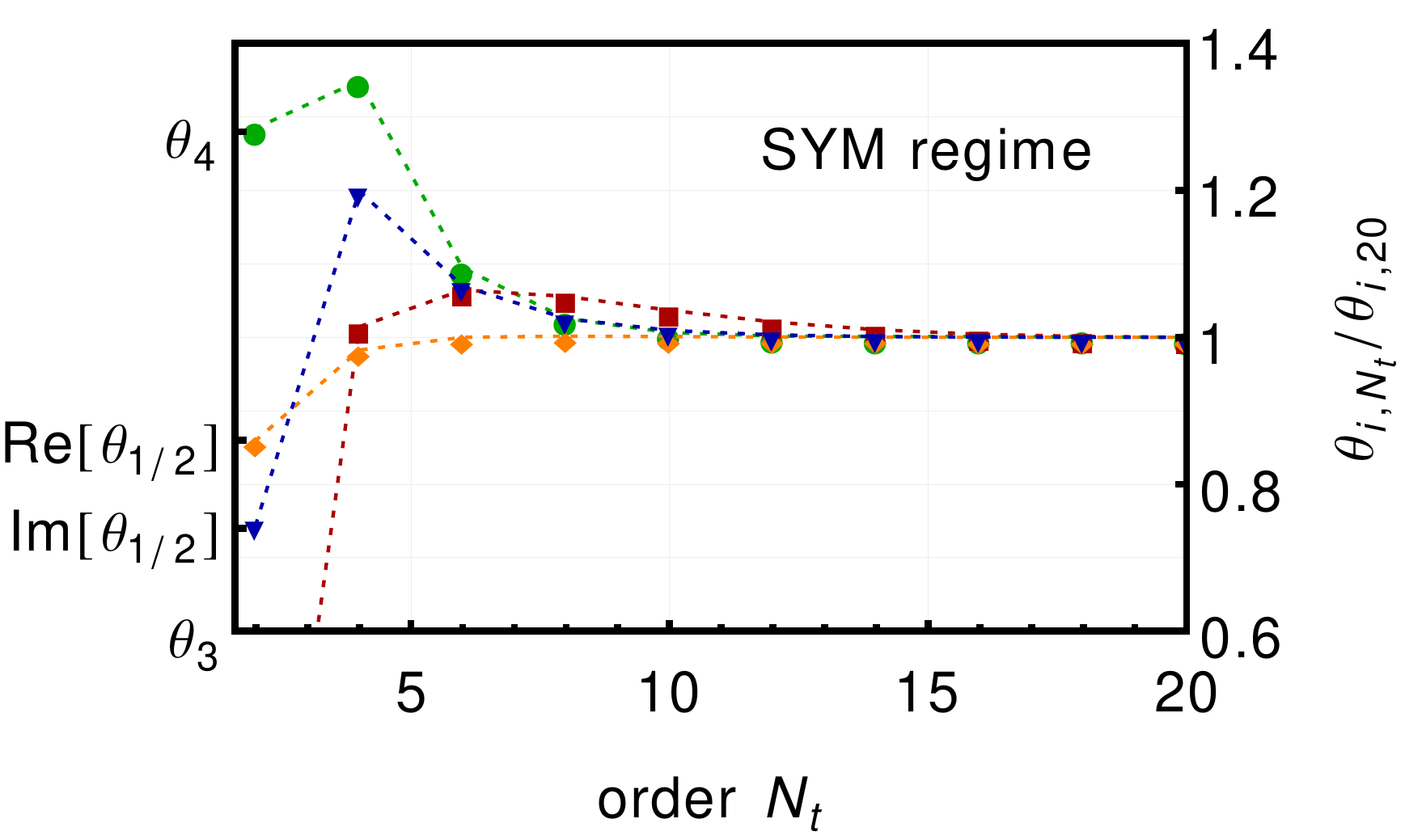}
	\hfill
	\includegraphics[width=0.46\textwidth]{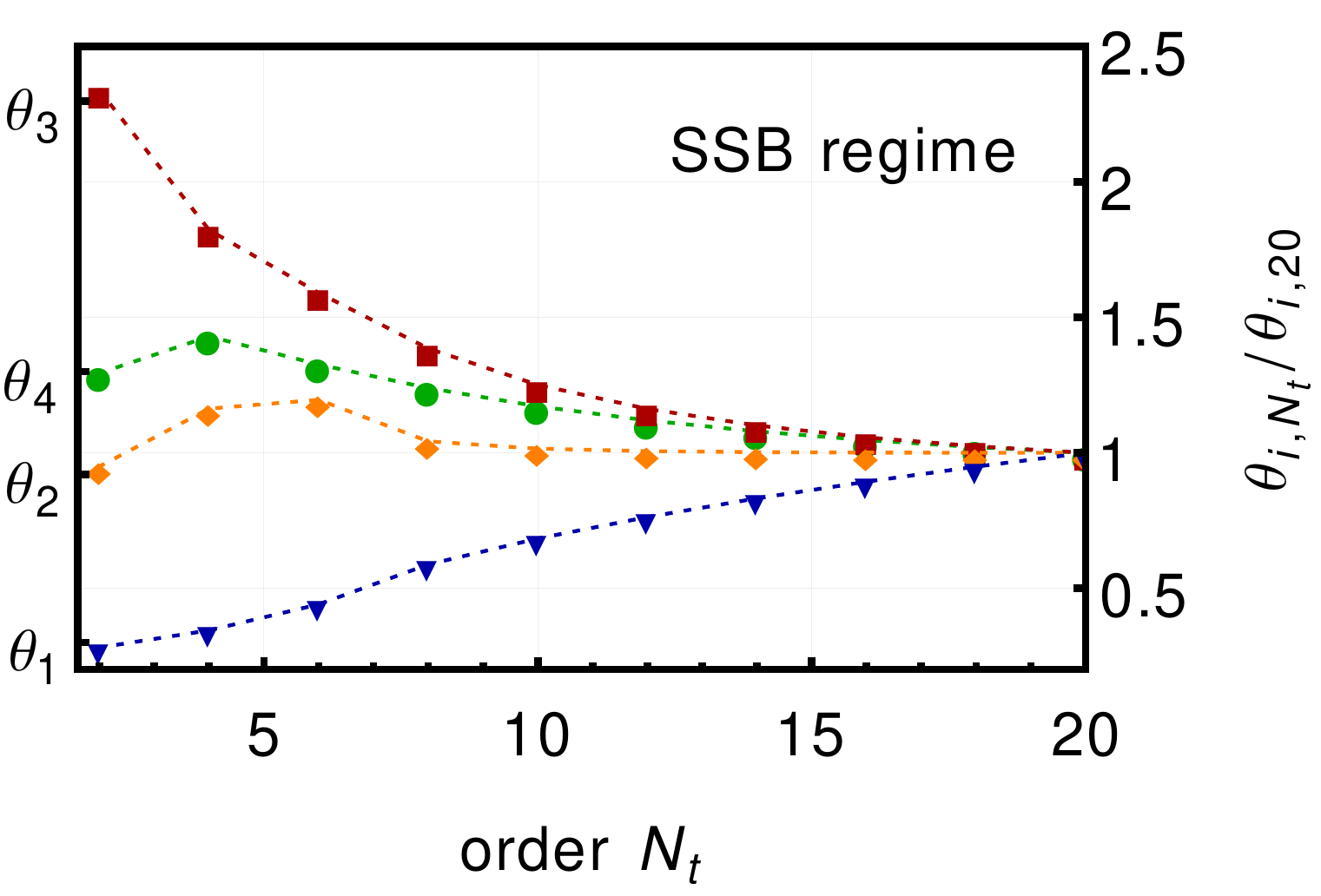}
\caption{
\label{fig:convergenceCEs}
Critical exponents for $N_f=N=1$ with increasing truncation order $N_t$ in the symmetric regime (left-hand panel) and in the regime of spontaneously broken symmetry (right-hand panel). All fixed point values are normalized to the values at the highest order.
}
\end{figure}
\begin{table}
\begin{tabular}{|c|c|c|c|c||c|c|c|c|}
$m_{\psi}^{\ast}$& $\kappa^{\ast}$& $\lambda_4^{\ast}$& $\lambda_{h\psi}^{\ast}$& $\eta_{\phi}$& $\theta_1$&$\theta_2$&$\theta_3$&$\theta_4$\\ \hline\hline
18.3 & $2\cdot 10^{-3}$& 129.4&-9497&0.50& 12.54 & 3.71& 1.71 & -1.90
\end{tabular}
\caption{\label{tab:SSB}Fixed-point coordinates of the masses and quartic couplings, the four leading critical exponents and the scalar anomalous dimension in $\rm LPA_{20}^{'}$ in the symmetry-broken parameterization.}
\end{table}
\begin{table}
\begin{tabular}{|c|c|c|c||c|c|c|c|}
$m_{\psi}^{\ast}$& $m_{\phi}^{2\,\ast}$& $\lambda_4^{\ast}$& $\lambda_{h\psi}^{\ast}$& $\theta_1$&$\theta_2$&$\theta_3$&$\theta_4$\\ \hline\hline
4.61 & -0.45& 25.3&-438& 2.2 + i\,2.1 & 2.2 - i\,2.1& 0.56 & -4.70
\end{tabular}
\caption{\label{tab:SYM}Fixed-point coordinates of the masses and quartic couplings and the four leading critical exponents in $\rm LPA_{20}$ in the symmetric parameterization.}
\end{table}
We now analyze the tentative fixed point in the Higgs-portal model in more detail, and in particular investigate its robustness under improvements of the approximation, i.e., enlargements of the truncation.
The fixed point features three relevant directions, cf.~Tab.~\ref{tab:SSB} and \ref{tab:SYM} as well as Fig.~\ref{fig:convergenceFPs} and \ref{fig:convergenceCEs}. In the symmetric parameterization, its properties appear reasonably stable under an increase of the truncation order in the scalar potential. 
We find that the fixed-point value for the scalar mass squared is negative. This indicates that the asymptotically safe fixed point lies in the symmetry-broken regime. Hence, one would expect an ansatz for the scalar potential in the symmetry-broken regime to be better suited to quantitatively describe the fixed-point regime. In particular, one would expect that lower orders in the truncation already provide better estimates for the critical exponents in the symmetry-broken regime. Indeed we rediscover the fixed point in the symmetry-broken parameterization given by Eq.~\eqref{eq:potentials}. However we find that the critical exponents at low orders of the truncation deviate significantly more from those at higher orders than in the symmetric parameterization, cf.~Fig.~\ref{fig:convergenceCEs}. This might be a consequence of the fact that a symmetry-broken parameterization is not well-suited to capture the properties of the fermionic sector: While the scalar potential features a nontrivial minimum in the symmetry-broken regime, the same need not be true for the Higgs-portal potential $V(\rho)$.\\
Canonically, the model features two relevant couplings, namely $m_{\phi}^2$ and $m_{\psi}$, with mass-dimension 2 and 1, respectively. Further, the quartic scalar self-interaction is marginally irrelevant and the Higgs-portal coupling has mass-dimension -1. The four largest critical exponents show deviations $\mathcal{O}(1)$ from the canonical values $2,1,0,-1$. Nevertheless, our truncation strategy of neglecting canonically higher-order couplings appears justified, as extending the truncation by canonically irrelevant couplings, such as, e.g., higher powers in the Higgs potential, only adds irrelevant directions. 

The three relevant and first irrelevant directions are superpositions of four quantities, namely the two masses, $m_{\psi},\, m_{\phi}$ and the two quartic couplings $\lambda_4, \, \lambda_{h\psi}$. In an effective field theory description of fermionic dark matter coupled via a Higgs-portal all  four couplings are free parameters that have to be constrained by experiment in order to test the model. The asymptotically safe UV completion of the model results in one relation among those four couplings. Thus it provides  a prediction for the irrelevant  superposition of couplings in terms of the relevant ones, cf. Sec.~\ref{sec:as_and_pred}. In particular, we can choose the IR values of two couplings in the scalar potential, i.e., the Higgs mass and quartic coupling. In a more realistic model these two parameters would already be fixed by experimental data. Asymptotic safety enforces a relation $\lambda_{h\psi}(m_\psi,m_{\phi,\text{exp}},\lambda_{4,\text{exp}}) = \lambda_{h\psi}(m_\psi)$ between the remaining two parameters $\lambda_{h\psi}$ and $m_\psi$.

\subsection{Predictivity and the Higgs-portal coupling}
\begin{figure}
	\centering
	\includegraphics[width=0.8\columnwidth]{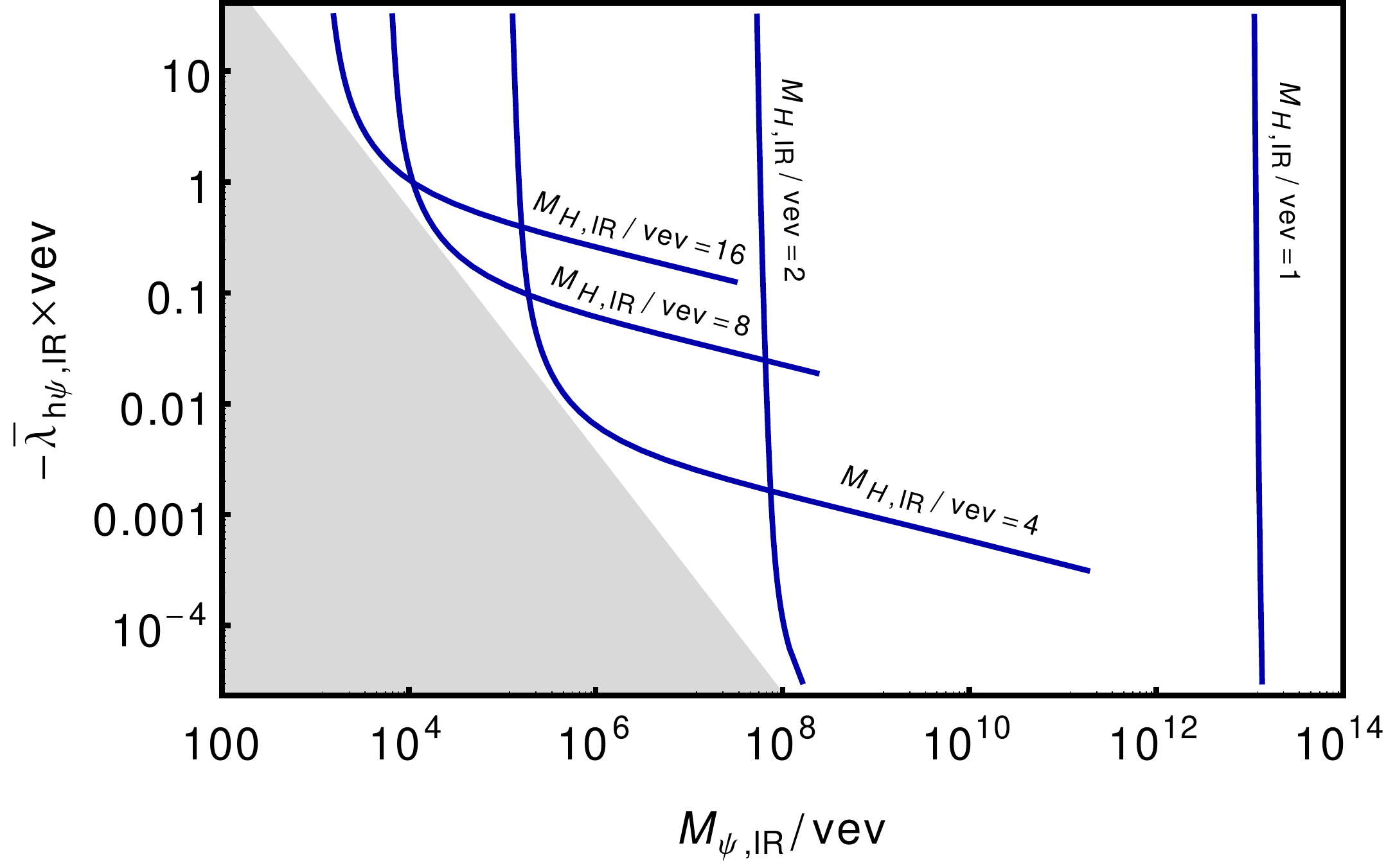}
\caption{
\label{fig:as-line}
Asymptotically safe prediction for the relation between Higgs-portal coupling $\bar\lambda_{h\psi,\text{IR}}\text{vev}$ and dark fermion mass $M_{\psi,\text{IR}}/\text{vev}$ in units of the vacuum expectation value, fixed to $\text{vev}^2=0.1$ for different scalar masses $M_{H,\text{IR}}/\text{vev}$ and $N=1$. The regime at small coupling (gray-shaded region) and small dark fermion mass cannot be accessed in the given truncation.
}
\end{figure}
\begin{figure}
	\centering
	\includegraphics[width=0.485\columnwidth]{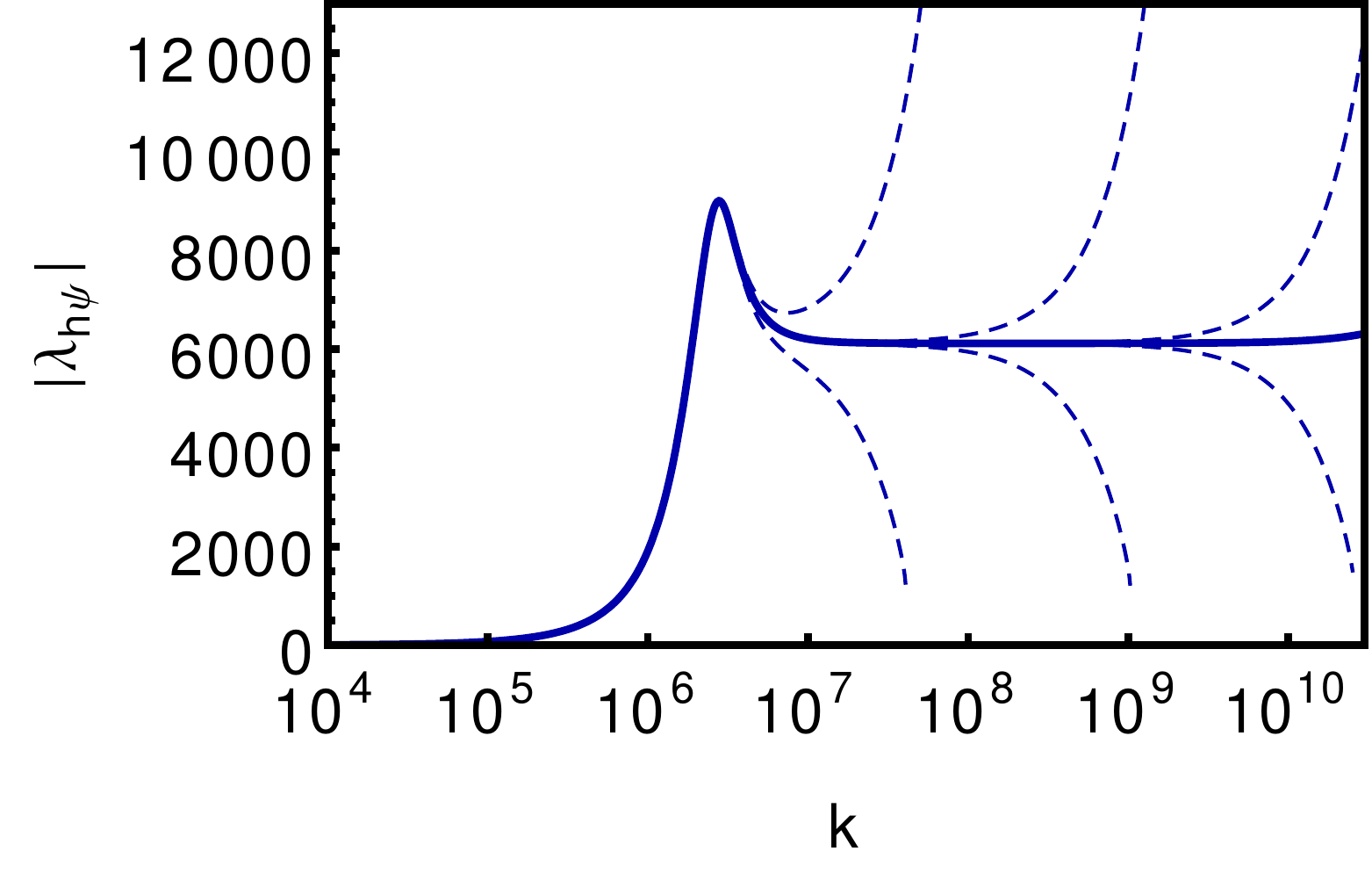}
	\hfill
	\includegraphics[width=0.495\columnwidth]{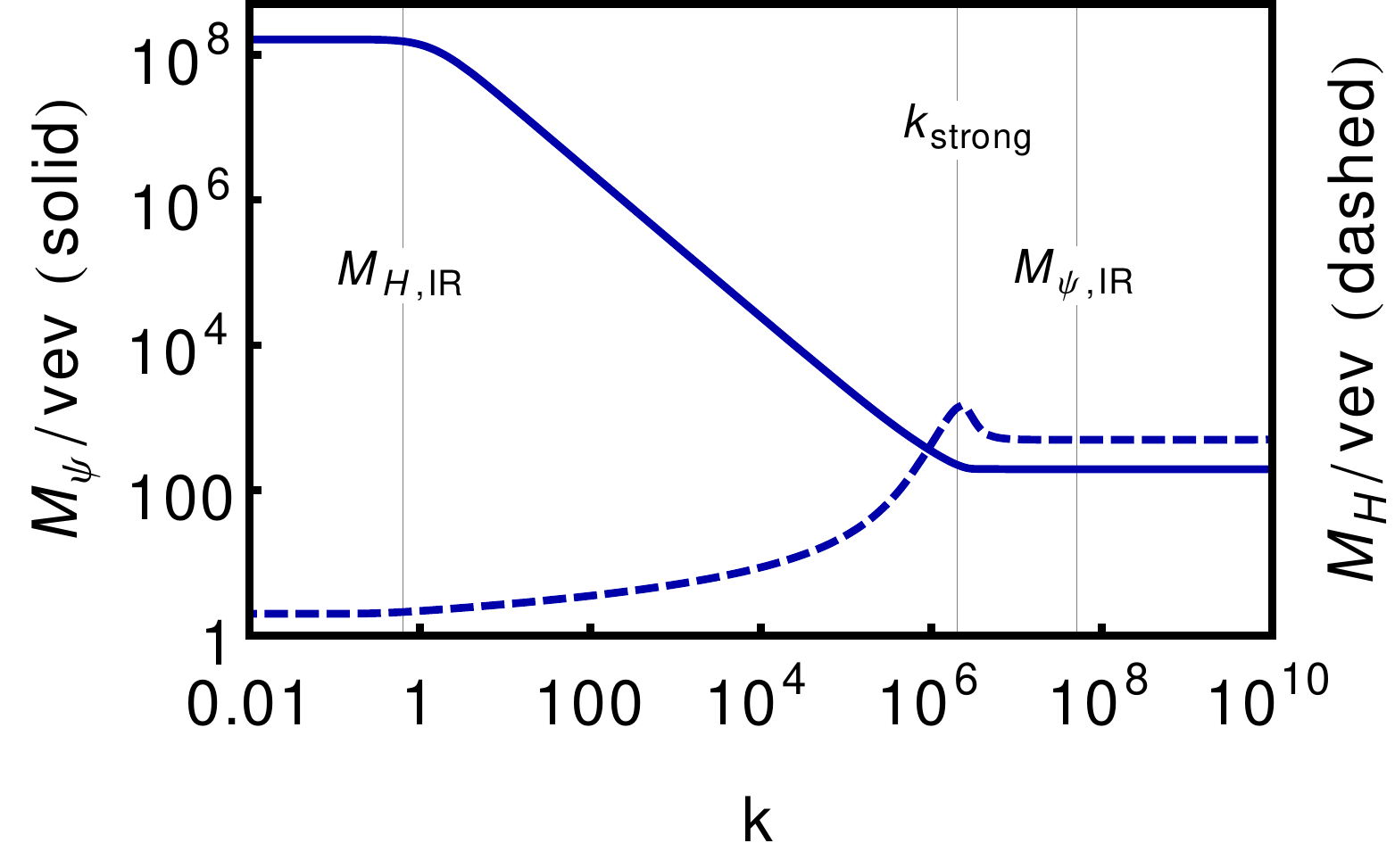}
\caption{
\label{fig:predictivity}
Left-hand panel: running of the dimensionless Higgs-portal coupling $\lambda_{h\psi}$ with RG scale $k$ from the far UV to the IR. All coupling values at scales above $M_{\psi,\text{IR}}/\text{vev}$ are attracted to trajectories close to the fixed point. This predicts $\lambda_{h\psi,\text{IR}}$. The values of the relevant couplings are fixed by $\text{vev}^2 = \kappa_\text{IR}=0.1$, $M_{H,\text{IR}}/\text{vev}=2$ and $M_{\psi,\text{IR}}/\text{vev}=1.6\times 10^{8}$.
\\
Right-hand panel: Corresponding running of the physical masses $M_{H}/\text{vev}(k)$ (dashed) and $M_{\psi}/\text{vev}(k)$ (solid). The plot shows UV scaling in a regime with dynamical bosons and fermions above $M_{\psi,\text{IR}}$. When the dark fermions become non-dynamical the flow departs from the fixed point. Below $M_{H,\text{IR}}$, after all degrees of freedom are decoupled, the flow enters the IR-scaling regime.
}
\end{figure}

To explicitly demonstrate the predictive power of asymptotic safety we construct RG flows from the UV to the IR. Along irrelevant directions such trajectories are quickly attracted to the critical hypersurface of the UV fixed point. Thus, fixing the IR vacuum expectation value $\kappa_\text{IR}$ and the IR effective masses
\be
M_{H,\text{IR}}= \sqrt{2\bar{\kappa}\, \lambda_4}\Big|_{k\rightarrow 0},\, \,  M_{\psi,\text{IR}} = 
\bar{m}_{\psi}\Big|_{k\rightarrow 0}, 
\ee
determines the deviation from the fixed point encoded in the values of $\kappa, \lambda_4$ and $m_{\psi}$. For all asymptotically safe trajectories, $\lambda_{h\psi}$ is then determined in terms of $\kappa, \lambda_4$ and $m_{\psi}$. For our specific example, we focus on $N_f=1$ and $N=1$ and evaluate the RG flow in the symmetry-broken phase in LPA 4 for simplicity. For $N=1$ there are no massless Goldstone modes, just as in the Standard Model. Thus, at
an infrared RG scale $k_{\rm IR}$ below the values of the physical masses $M_{H,\text{IR}}$ and $M_{\psi,\text{IR}}$, the threshold effects in Eqs.~\eqref{eq:beta_portal_SYM}-\eqref{eq:beta_mPhi_SYM}, and their symmetry-broken counterparts in Eq.~\eqref{eq:runningCouplingsAndPotential} lead to an automatic decoupling of the corresponding degrees of freedom,  cf.~right panel in Fig.~\ref{fig:predictivity}. In this regime, all dimensionful quantities stop running, whereas their dimensionless counterparts scale canonically.
We can thus read off the IR values of all couplings once $k$ has dropped below $k_{\rm IR}$.
\\
More specifically, the system is determined by the interplay of the two mass scales $M_{H,\text{IR}}$ and $M_{\psi,\text{IR}}$. The light scalar mass $M_{H,\text{IR}}$ sets the IR-scale at which the scalar degree of freedom decouples. Correspondingly, the heavier fermion mass $M_{\psi,\text{IR}}$ sets the decoupling scale of the dark matter fermions.
The UV scaling regime relies on a strongly coupled fermion sector to balance the classical scaling of $\lambda_{h\psi}$, cf.~Sec.~\ref{sec:higgs-portal}. Hence, the freeze-out of the fermionic degrees of freedom at $M_{\psi,\text{IR}}$ also determines the scale at which trajectories depart from the scaling regime at the UV fixed point. The fermion mass thus sets the dynamically emergent transition scale from the fixed-point regime.
\\
Towards the IR, trajectories with a scaling-regime in the IR are first pulled towards increasing $\lambda_{h\psi}$. Between the two scales $M_{H,\text{IR}}$ and $M_{\psi,\text{IR}}$, the system thus develops an additional dynamical scale $k_\text{strong}$ at which bosonic fluctuations in the running of $\lambda_{h\psi}$ and $\lambda_4$ outgrow the fermionic contributions, cf.~Fig.~\ref{fig:predictivity}. The fermionic couplings, which are dominant above $k_\text{strong}$, tend to drive the system towards stronger couplings. Below $k_\text{strong}$, the scalar contributions push the system back into a more weakly coupled regime.
\\
In Fig.~\ref{fig:predictivity} we demonstrate how the model dynamically determines $\bar{\lambda}_{h\psi}(k_{\rm IR})$: We choose UV initial-conditions for the relevant directions such that $\text{vev}^2 = \kappa_\text{IR}=0.1$, $M_{H,\text{IR}}/\text{vev}=1$ and $M_{\psi,\text{IR}}/\text{vev}=1.9\times 10^{11}$. The left panel in Fig.~\ref{fig:predictivity} highlights that the prediction of $\lambda_{h\psi}$ is independent of the UV initial conditions, as long as the UV cutoff scale is chosen high enough, since a range of UV-values is attracted towards the critical trajectory. Effectively, the asymptotically safe regime washes out physics determining $\lambda_{h\psi}$ above $M_{\psi,\text{IR}}/\text{vev}$. In particular, setting the deep-UV value of the Higgs-portal coupling to the fixed-point value, the IR value is unique. Fig.~\ref{fig:predictivity} clearly shows that even tiny deviations from the critical trajectory in the deep IR are incompatible with asymptotic safety in the UV.
\\
Further, we scan through a range of UV values of $m_{\psi}$, translating into an IR range of dark-fermion masses. For each triple $(\kappa, \lambda_4, m_{\psi})$, $\lambda_{h\psi}$ is fixed uniquely by asymptotic safety. The asymptotic safety paradigm then predicts a relation $\bar{\lambda}_{h\psi}(M_\psi)$ for the dark-matter candidate. This ``asymptotic-safety line'' is shown in Fig.~\ref{fig:as-line}. Values of the mass and portal coupling that deviate from this line will result in divergences at higher energies: the model becomes non-fundamental, i.e., reduces to the less predictive effective field theory setup.
\\
If the property of asymptotic safety persisted under the inclusion of further Standard Model degrees of freedom, such a relation could be exploited to guide experimental searches for dark matter: Demanding that the dark fermion constitutes the complete dark-matter relic-density leads to a relation $\bar{\lambda}_{h\psi}(M_{\psi})$.  If a fixed point were indeed available, then demanding asymptotic safety in the model  would lead to a second, \emph{independent} relation between these two parameters (if all other parameters such as, e.g., the Higgs mass, are held fixed). Let us assume that this is indeed the case in order to briefly discuss the phenomenological consequences.
There are two distinct possibilities for these two relations: If they cannot be imposed simultaneously, asymptotic safety in this model is incompatible with the dark fermion constituting all of the observed dark matter. Depending on the location of the asymptotic-safety relation $\lambda_{h\psi}(m_{\psi})$, it might not be possible to accommodate any of the observed dark matter, if the asymptotic-safety line lies in the regime where an over-density is produced. The second possibility results in one unique point where the relic-density line and the asymptotic-safety line intersect, leading to a unique prediction for $m_{\psi}$ and $\lambda_{h\psi}$ under the combined assumptions of 100$\%$ relic density and asymptotic safety.  Given that for $m_{\psi}\gtrsim m_{\rm Higgs}$, the relic-density line in the $\bar{\lambda}_{h\psi}-M_{\psi}$ plane asymptotes to a constant, see \cite{Beniwal:2015sdl}, our explicit results for the asymptotic safety prediction would indeed imply a single intersection point. We stress that within our toy model such inferences on experimental tests cannot be made. However our model does highlight the  potential predictive power of the asymptotic-safety paradigm.

\section{Conclusions and outlook}{\label{sec:conclusions}}
We explore the Higgs-portal coupling to dark fermions and discover potential hints for asymptotic safety in this model. The symmetries of the model restrict the interaction to a quartic coupling between two powers of the Higgs field and two fermions, instead of a Yukawa-type coupling with one power of the Higgs field. While the latter is perturbatively renormalizable in four dimensions, the former is not. Within canonical power counting, the Higgs portal model therefore is an effective field theory. Going beyond canonical power counting, quantum fluctuations can alter the scaling dimensions of couplings and could induce nontrivial fixed points of the RG, generalizing asymptotic freedom to asymptotic safety which underlies a notion of nonperturbative renormalizability. We find hints for asymptotic safety in the Higgs-portal model. An interacting fixed point exists in our approximation.  If confirmed beyond our truncation, our findings would imply that scalars in four dimensions are no longer trivial  if coupled to a fermionic sector. Our model circumvents the conclusion in \cite{Bond:2018oco}, as it relies on the presence of a perturbatively nonrenormalizable Higgs-portal coupling.
\\
Asymptotic safety in the UV could have intriguing consequences for the IR: In our approximation, all UV complete RG trajectories that emanate from the fixed point impose a relation among the couplings. In particular, given the Higgs mass, fermion mass and Higgs quartic selfinteraction, the Higgs-portal coupling might no longer be a free parameter. Instead, its value can be calculated given the other three quantities. In this paper, we demonstrate this explicitly within our toy model.
\\
Further, we analyze the RG flow away from the potential fixed point with a view towards lower bounds on the Higgs mass arising from vacuum stability. Fermionic fluctuations coupled through a Yukawa coupling tend to increase the lower bound, generating a tension between a positive or vanishing Higgs-quartic interaction up to the Planck scale and a Higgs mass of 125 GeV.
Intriguingly, fermionic fluctuations coupled through the Higgs portal can have either an increasing or a decreasing effect on the lower bound on the Higgs mass, depending on the value of their mass. For small masses -- measured in units of the RG scale -- they lead to a decrease of the lower bound on the Higgs mass, just as scalar fluctuations coupled through a Higgs portal do.  Accordingly, fermionic fluctuations coupled through a Higgs portal might contribute to reducing the tension between a vanishing Higgs quartic coupling at the Planck scale and the measured Higgs mass of 125 GeV. The converse effect is realized for larger fermion masses.
\\
Beyond the potential phenomenological interest, our model is of intrinsic interest as it might constitute a simple example for asymptotic safety in four dimensions with a minimal set of degrees of freedom, namely one scalar and one fermion  -- extensions to $N_f>1$ and $N>1$ appear to also accommodate fixed points. Intriguingly, in our approximation, both mechanisms that can underlie asymptotic safety are at play here: To induce an interacting fixed point in the dimensionless quartic Higgs selfinteraction, the fermionic and bosonic fluctuations have to cancel. To induce an interacting fixed point in the dimensionful Higgs-portal coupling, the canonical scaling term balances against the contribution from quantum fluctuations. Regardless of whether asymptotic safety persists in the Higgs portal sector once further Standard Model degrees of freedom are included, this model could therefore be an appealingly simple but rich example of the mechanisms underlying asymptotic safety in other models.
To strengthen the indications for asymptotic safety in the model, truncations are required that take into account further effects in the strongly coupled fermionic sector, such as, e.g., four-fermion interactions, and four-fermion-two-scalar interactions.\\
We emphasize that indications for the existence of the fixed point arise from truncated RG flows, and we do not see any hints that the fixed point becomes completely perturbative in any of the limits $N_s \rightarrow \infty$($d=4$), $N_f\rightarrow 0$ ($d=4-\epsilon$), or $d \rightarrow 3$. Moreover, the deviations from canonical scaling in our truncation are of order 1, hinting at a nonperturbative nature of the fixed point. Accordingly, a lattice study of the system would be highly worthwhile, as it could shed further light on the existence of a nonperturbative fixed point.
\\
There are several intriguing questions that arise as a consequence of our results.
We have focused on the parity-conserving Higgs portal, and have set the parity violating interaction
\be
\mathcal{L}_{\rm HP\, non-parity} = \bar{\lambda}_{h\psi,\, np}\bar{\psi}\gamma_5\psi \phi^{\dagger}\phi,
\ee
to zero. Understanding the impact of parity on the fixed-point structure is of interest from a conceptual as well as a phenomenological point of view, since observational constraints on the two interaction channels differ, see, e.g., \cite{LopezHonorez:2012kv}.
\\
The fixed point in our truncation can be continued down to $d=3$, where it provides a new universality class, potentially governing the scaling regime in the vicinity of a continuous phase transition. It is an intriguing question whether an appropriate condensed-matter system exists. 
\\

\noindent\emph{Acknowledgements:}\\
We gratefully acknowledge insightful discussions with M.~M.~Scherer.
This research is supported by the DFG under the Emmy-Noether program, grant no.~Ei/1037-1. This research is also supported in part by Perimeter Institute for Theoretical Physics. Research at Perimeter Institute is supported
by the Government of Canada through Industry Canada and by the Province of Ontario through the Ministry of Economic Development and Innovation. A.~Held also acknowledges support by the Studienstiftung des deutschen Volkes.

\appendix

\section{Connection to perturbation theory in the scalar sub-sector}

The fixed point that is present in our truncation of the Higgs portal model exhibits deviations $\mathcal{O}(1)$ from canonical scaling in four dimensions, i.e., it appears to be nonperturbative in nature.  (While one might draw a similar conclusion from the fixed-point values, the couplings can of course be rescaled to make the fixed-point value arbitrarily small without changing the critical exponents.)
Establishing a connection to a perturbative fixed point would provide further evidence in favor of the existence of this fixed point beyond our truncation. To explore this question, we consider the limit $N_f \rightarrow 0$ in $d=4-\epsilon$, with $\epsilon$ lying in the interval between 0 and 1.  At finite $N$, the limiting case $N_f=0$ results in a purely scalar model.
Scalar $O(N)$-models are perturbatively renormalizable in four dimensions. Yet, they are trivial \cite{Aizenman:1981du,Frohlich:1982tw, Wolff:2009ke}, i.e., the free fixed point is infrared attractive in the coupling. In a perturbative expansion of the beta function, this leads to a positive one-loop coefficient.
In $d=4-\epsilon$ dimensions the canonical dimension contributes a negative term that balances against the positive one-loop coefficient. This results in a fixed point at $\lambda_4^{\ast} \sim \epsilon$. This perturbative fixed point can be extended to the Wilson-Fisher fixed point in three dimensions, see, e.g., \cite{Canet:2003qd,Litim:2010tt,Hasenbusch:2011yya,ElShowk:2012ht} for studies of the $N=1$ Ising model with various methods.
\\
The non-perturbative fixed point that we find when coupling the O($N$) model in four dimensions to $N_f$ fermions via a non-vanishing Higgs-portal-like coupling, is based on fermionic fluctuations that act as an effective antiscreening contribution. This contribution balances bosonic fluctuations in the quartic coupling and allows for a non-perturbative fixed point in the quartic coupling $\lambda_4$.
When fermionic fluctuations are switched off continuously, i.e., $N_f\rightarrow 0$, the scalar sector continuously approaches the Wilson-Fisher fixed point. We explicitly check this for $N=1,2$. While the limiting case $N_f=0$ must of course yield the purely scalar fixed point, the continuity of the limit $N_f \rightarrow 0$ is nontrivial. It shows that our fixed point can partially be understood as an extension of the Wilson-Fisher universality calss by a nontrivial fermionic sector.

\begin{figure}
	\includegraphics[width=0.475\columnwidth]{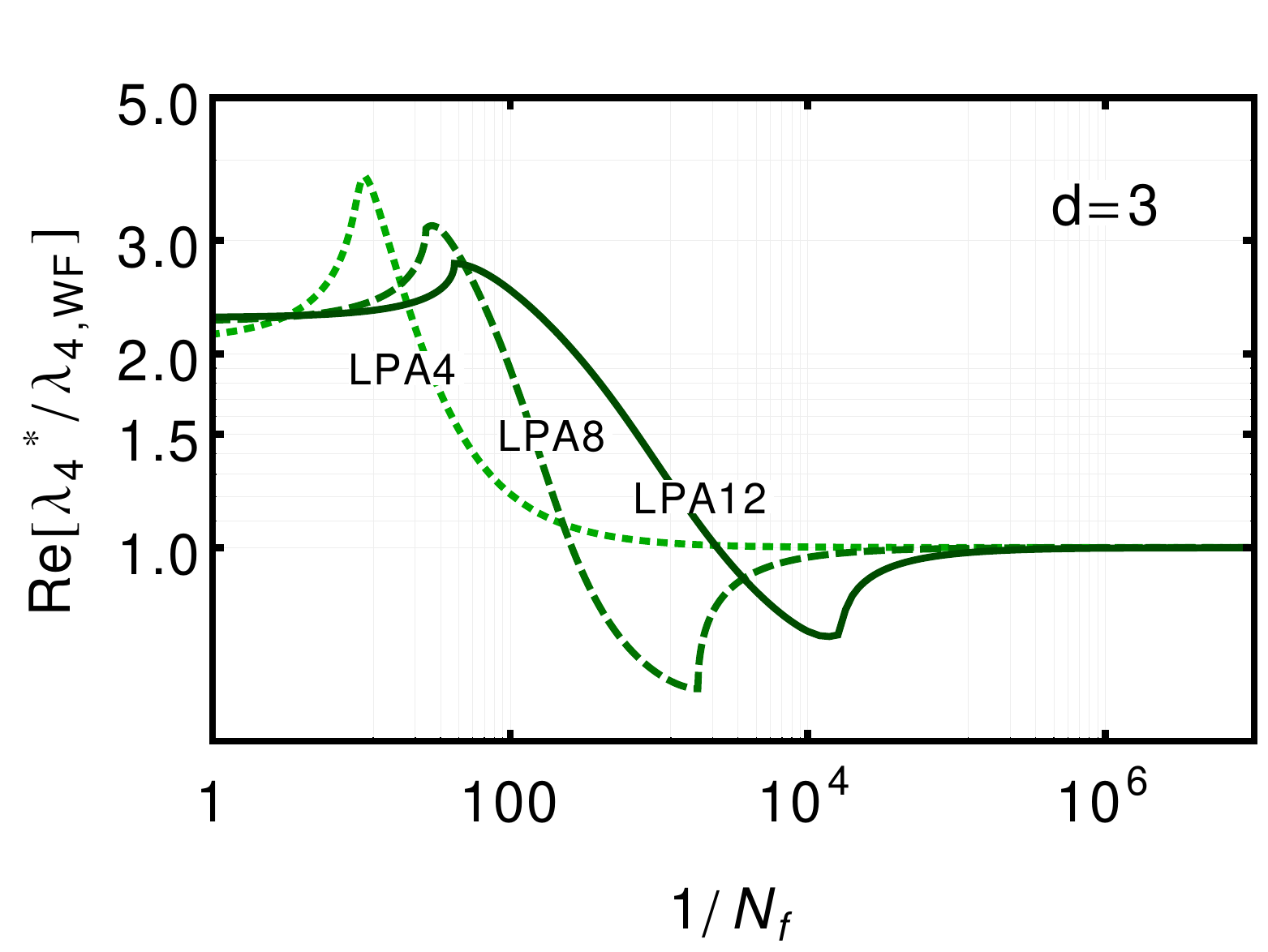}
	\hfill
	\includegraphics[width=0.515\columnwidth]{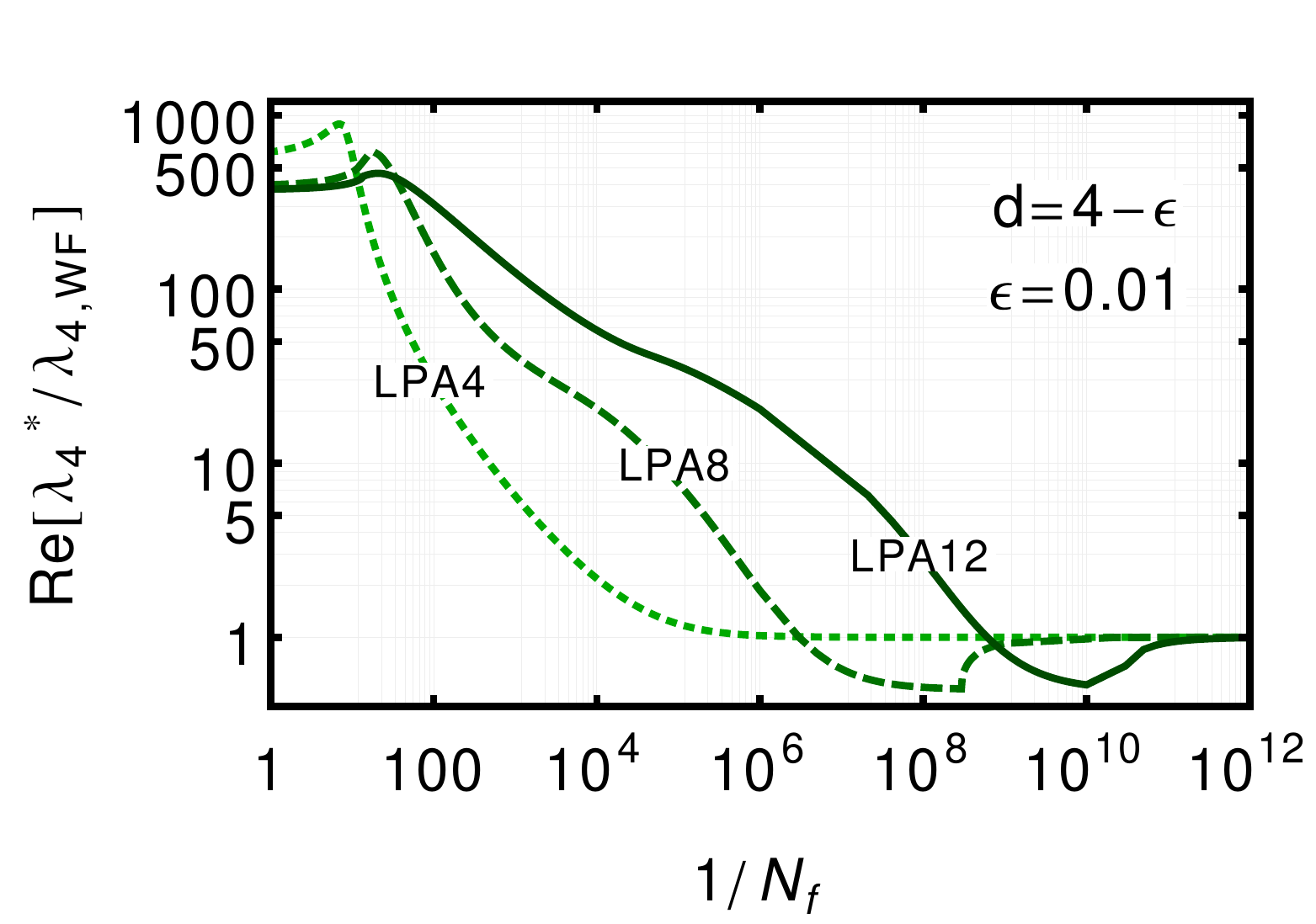}
\caption{
\label{fig:perturbativeScalarSectorIn4d}
Fixed point values of the scalar quartic coupling $\lambda_4$ for $N=2$ during a decoupling of the fermionic sector, i.e., $1/N_f\rightarrow\infty$. All fixed point values are normalized to values at the purely scalar Wilson-Fisher fixed point of the corresponding truncation. The left-hand panel shows convergence to the non-perturbative Wilson-Fisher fixed point in $d=3$. The right-hand panel shows convergence to perturbative fixed point values near $d=4$, i.e. $d=3.99$.
}
\end{figure}

In Fig.~\ref{fig:perturbativeScalarSectorIn4d} the behavior of $\lambda_4$ is explicitly shown in several orders of the truncation ($\rm LPA_4$, $\rm LPA_8$ and $\rm LPA_{12}$ at $N=2$). The deformation is continuous in $\rm LPA_4$, while higher truncation orders seem to cause the fixed point to divert into the complex plane at intermediate $1/N_f$. At all orders it becomes real again for large 1/$N_f$ and then converges to the Wilson-Fisher fixed point of the purely scalar sector.
In $d=3$, (cf.~left-hand panel in Fig.~\ref{fig:perturbativeScalarSectorIn4d}) the fixed-point values continuously approach those of the Wilson-Fisher fixed point in the corresponding truncation of the functional RG flow, see e.g. \cite{Litim:2002cf}.
In $d=4+\epsilon$ (cf.~right-hand panel in Fig.~\ref{fig:perturbativeScalarSectorIn4d}) we recover a perturbatively controlled limit for a subsector of the fixed point in our model. While the fermions decouple, $\lambda_{h\psi}$ and $m_{\psi}$ approach finite asymptotic values.
The fixed point in the scalar sector continuously approaches a fixed-point value $\lambda_4^{\ast}\sim \epsilon$ and accordingly merges with the Gaussian fixed point in four dimensions. For $N=1$ this smooth connection to the Wilson-Fisher fixed point appears to be obstructed by a divergence in the fixed-point values.
\\
Within our truncation the fermionic sector of the model seems genuinely non-perturbative. In four dimensions the classical scaling term is of $\mathcal{O}(1)$ due to the negative mass-dimension of $\lambda_{h\psi}$. Balancing this term requires a non-perturbative regime in the couplings $m_\psi$ and $\lambda_{h\psi}$. In $d=3+\epsilon$ the classical scaling of $\lambda_{h\psi}$ becomes $\mathcal{O}(\epsilon)$ and there is no longer a need for non-perturbative effects in the fermionic sector of the model. On the other hand, now the scalar sector, in particular $\lambda_4$ which has a classical scaling of $1-\epsilon$ in $d=3+\epsilon$, requires non-perturbative dynamics. The limit of non-dynamical bosons in $d=3$ on the other hand leads to the vanishing of all quantum fluctuations in our truncation, simply because there are no purely fermionic interactions. In summary, there does not appear to be a fully perturbative limit of the fixed point that we have discovered, as only its scalar sector can be understood as a fermion-induced deformation of a perturbative fixed point in $d=4-\epsilon$.

\section{Threshold integrals}
\label{app:thresholdIntegrals}
The explicit threshold functions integrated with a Litim-type regulator  read
\begin{align}
\label{eq:thresholdIntegrals}
	l_n^{(B)d}(\omega_B;\eta_B) &=
		\frac{4v_d}{d}
		\frac{1}{(1+\omega_B)^{n+1}}	
		\left(1-\frac{\eta_B}{d+2}\right)\;,
	\\[2ex]
	l_n^{(F)d}(\omega_F;\eta_F) &=
		\frac{4v_d}{d}
		\frac{1}{(1+\omega_F)^n}	
		\left(1-\frac{\eta_F}{d+1}\right)\;,
	\notag\\[2ex]
	l_{n_B,n_F}^{(BF)d}(\omega_B,\omega_F;\eta_B,\eta_F) &=
		\frac{4v_d}{d}
		\frac{1}{(1+\omega_B)^{n_B}(1+\omega_F)^{n_F}}
		\left(
			\frac{n_B\left(1-\frac{\eta_B}{d+2}\right)}{1+\omega_B}
			+\frac{n_F\left(1-\frac{\eta_F}{d+1}\right)}{1+\omega_F}	
		\right)
	\notag\;,
\end{align}
where $v_d$ is the surface of a d-dimensional sphere, i.e.,
\begin{align}
	v_d = \frac{1}{2^{d+1}\pi^{d/2}\Gamma(d/2)}\;.
\end{align}

\section{Projection rules}
\label{sec:projectionRules}
The following projection rules for the fermionic couplings $m_\psi^2$, $\lambda_{h\psi}$ and the potential hold:
\begin{align}
	\beta_{m_\psi} &= 
	(\eta_\psi-1)m_\psi 
	- i\frac{\delta}{\delta(\bar\psi\psi)}\dot\Gamma_k\Bigg|_{\substack{(\bar\psi\psi)\rightarrow 0\\\rho\rightarrow\kappa }}
	+ \lambda_{h\psi}\beta_\kappa + (d-2+\eta_\phi)\lambda_{h\psi}\kappa
	\;,\notag\\
	\beta_{\lambda_{h\psi}} &= (d-3+\eta_\phi+\eta_\psi)\lambda_{h\psi}
	- i\frac{\delta}{\delta \rho}\frac{\delta}{\delta(\bar\psi\psi)}\dot\Gamma_k\Bigg|_{\substack{(\bar\psi\psi)\rightarrow 0\\\rho\rightarrow\kappa }}
	\;,\notag\\
	\dot{U} &= -d\,U(\rho) + (d-2+\eta_\phi)\rho\,U'(\rho)
	+\dot\Gamma_k\Bigg|_{(\bar\psi\psi)\rightarrow 0}\;,
	\label{eq:projections}
\end{align}
where, in the functional derivatives, the fields are treated as constant external fields.
Plugging in the ansatz for the effective action $\Gamma_k$ (cf. Eq.~\eqref{eq:effectiveAction}) and using Litim-type threshold integrals (cf. Eq.~\eqref{eq:thresholdIntegrals}) gives the explicit form of the $\beta$-functions, cf. Eq.~\eqref{sec:explicitGeneralBetas}.
We then use a polynomial ansatz for the potential $U(\rho)$ in both the symmetric and symmetry-broken regime, cf. Eq.~\ref{eq:potentials}.
The projections on single couplings of such an expansion are given by
\begin{align}
	\label{eq:potentialProjections}
	\text{SYM:}\quad
	\beta_{m_\phi} &= \dot{U}'(\rho)\Big|_{\rho\rightarrow 0}
	\;,\notag\\
	\beta_{\lambda_{2n}} &= \dot{U}^{(n)}(\rho)\Big|_{\rho\rightarrow 0}\;\; \forall\;\;N>n>1
	\;,\\
	\text{SSB:}\quad
	\beta_{\kappa} &= -\frac{\dot{U}'(\rho)}{U''(\rho)}\Big|_{\rho\rightarrow 0}
	\;,\notag\\
	\beta_{\lambda_{2n}} &= \dot{U}^{(n)}(\rho)\Big|_{\rho\rightarrow 0}
	+\lambda_{2(n+1)}\,\beta_\kappa
	\;\; \forall\;\;N-1>n>1
	\;,\notag\\
	\beta_{\lambda_{2n}} &= \dot{U}^{(n)}(\rho)\Big|_{\rho\rightarrow 0}
	\;\; \text{for}\;\;n=N
	\;.
\end{align}

\section{$\beta$-functions for general $N_f$ and $N$}
\label{sec:explicitGeneralBetas}
The explicit running of the scalar potential and the fermionic couplings is given by
\begin{align}
	\dot{U}(\rho) &=
	\left(d+\eta _{\phi }-2\right) U'(\rho )\,\rho  - d U(\rho )	
	+c_d \left(-\frac{d_{\gamma } N_f \left(1-\frac{\eta _{\psi }}{d+1}\right)}{1+\omega_\psi(\rho)}+\frac{\left(N_s-1\right) \left(1-\frac{\eta _{\phi }}{d+2}\right)}{1+\omega_G(\rho)}+\frac{1-\frac{\eta _{\phi }}{d+2}}{1+\omega_\rho(\rho)}\right)\;,
   \\
	\beta_{m_\psi} &= 
	\left(-1+\eta _{\psi }\right) m_{\psi }
	+\lambda_{h\psi}\partial_t\kappa + (d-2+\eta_\phi)\lambda_{h\psi}\kappa
	+c_d \Bigg[
	-\lambda _{\text{h$\psi $}} \left(\frac{\left(N_s-1\right) \left(1-\frac{\eta _{\phi
   }}{d+2}\right)}{\left(1+\omega_G\right){}^2}+\frac{1-\frac{\eta _{\phi }}{d+2}}{\left(1+\omega_\rho\right){}^2}\right)
	\notag\\&\quad\quad\quad\quad\quad\quad\quad
	+4 \kappa  \lambda _{\text{h$\psi $}}^2 \sqrt{\omega_\psi} \left(\frac{1-\frac{\eta _{\psi }}{d+1}}{\left(1+\omega_\rho\right) \left(1+\omega_\psi\right){}^2}+\frac{1-\frac{\eta _{\phi}}{d+2}}{\left(1+\omega_\rho\right){}^2 \left(1+\omega_\psi\right)}\right)
   \Bigg]\;,
   \\
   \beta_{\lambda_{h\psi}} &=
   \left(d+\eta _{\psi }+\eta _{\phi }-3\right)\lambda _{\text{h$\psi $}} 
   +4c_d\Bigg[
   \kappa  \lambda _{\text{h$\psi $}}^3 \left(\frac{1-\frac{\eta _{\psi }}{d+1}}{\left(1+\omega_\rho\right) \left(1+\omega_\psi\right){}^2}+\frac{1-\frac{\eta _{\phi }}{d+2}}{\left(1+\omega_\rho\right){}^2 \left(1+\omega_\psi\right)}\right)
   \notag\\
   &
   -2 \kappa  \lambda _{\text{h$\psi $}}^3 \omega_\psi \left(\frac{2 \left(1-\frac{\eta _{\psi }}{d+1}\right)}{\left(1+\omega_\rho\right) \left(1+\omega_\psi\right){}^3}+\frac{1-\frac{\eta _{\phi
   }}{d+2}}{\left(1+\omega_\rho\right){}^2 \left(1+\omega_\psi\right){}^2}\right)\notag\\
   &
   -\kappa  \lambda _{\text{h$\psi $}}^2 \sqrt{\omega_\psi} \left(\frac{\omega_\rho' \left(1-\frac{\eta _{\psi }}{d+1}\right)}{\left(1+\omega_\rho\right){}^2 \left(1+\omega_\psi\right){}^2}+\frac{2 \omega_\rho' \left(1-\frac{\eta _{\phi }}{d+2}\right)}{\left(1+\omega_\rho\right){}^3 \left(1+\omega_\psi\right)}\right)
   \notag\\
   &
   +\lambda _{\text{h$\psi $}}^2 \sqrt{\omega_\psi} \left(\frac{1-\frac{\eta _{\psi }}{d+1}}{\left(1+\omega_\rho\right) \left(1+\omega_\psi\right){}^2}+\frac{1-\frac{\eta _{\phi }}{d+2}}{\left(1+\omega_{\rho
   }\right){}^2 \left(1+\omega_\psi\right)}\right)\notag\\
   &
   +\frac{\lambda _{\text{h$\psi $}}}{2} \left(\frac{\omega_G' \left(N_s-1\right) \left(1-\frac{\eta _{\phi }}{d+2}\right)}{\left(1+\omega_G\right){}^3}+\frac{\omega_\rho'\,\left(1-\frac{\eta _{\phi
   }}{d+2}\right)}{\left(1+\omega_\rho\right){}^3}\right)
   \Bigg]\;,
	\label{eq:runningCouplingsAndPotential}
\end{align}
where the squared effective masses $\omega_\rho(\rho)$, $\omega_G(\rho)$ and $\omega_\psi(\rho)$ are given in Eq.~\eqref{eq:effectiveMasses}, and we use the shorthand notation $\omega_i = \omega_i(\kappa)$ and $\omega_i' = \frac{\partial\omega_i(\rho)}{\partial\rho}\Big|_{\rho=\kappa}$.

\end{document}